\documentclass[fleqn,usenatbib]{emulateapj}
\usepackage{graphicx}
\usepackage{natbib}
\usepackage{epstopdf}
\usepackage{amssymb}
\usepackage{array, makecell}
\usepackage{amsmath}
\usepackage{mathptmx}
\usepackage{booktabs}
\usepackage{hyperref}
\usepackage{float}
\usepackage[normalem]{ulem}
\usepackage{color}

\DeclareRobustCommand{\VAN}[3]{#2}
\let\VANthebibliography\thebibliography
\def\thebibliography{\DeclareRobustCommand{\VAN}[3]{##3}\VANthebibliography}
\slugcomment{draft v2}

\shorttitle{Response of BH gaps to external changes}
\shortauthors{Kisaka et al.}

\begin{document}


\title{The response of black hole spark gaps to external changes: A production mechanism of rapid TeV flares?}


\author{Shota Kisaka\altaffilmark{1}}
\email{kisaka@hiroshima-u.ac.jp}
\author{Amir Levinson\altaffilmark{2}}
\author{Kenji Toma\altaffilmark{3,4}}
\author{Idan Niv\altaffilmark{2}}


\altaffiltext{1}{Department of Physics, Hiroshima University, Higashi-Hiroshima, 739-8526, Japan}
\altaffiltext{2}{School of Physics \& Astronomy, Tel Aviv University, Tel Aviv 69978, Israel}
\altaffiltext{3}{Frontier Research Institute for Interdisciplinary Sciences, Tohoku University, Sendai, 980-8578, Japan}
\altaffiltext{4}{Astronomical Institute, Graduate School of Science, Tohoku University, Sendai, 980-8578, Japan}


\begin{abstract}
We study the response of a starved Kerr black hole magnetosphere to abrupt changes in the intensity of 
disk emission and in the global magnetospheric current, by means of 1D general relativistic particle-in-cell simulations.
Such changes likely arise from the intermittency of the accretion process. 
We find that in cases where the pair production opacity contributed by the soft disk photons is modest, as in, e.g., M87,
such changes can give rise to delayed, strong TeV flares, dominated by curvature emission of particles accelerated in the gap. 
The flare rise time, and the delay between the external variation and the onset
of the flare emitted from the outer gap boundary,
are of the order of the light crossing time of the gap.  The rapid, large amplitude TeV flares observed in M87 and, perhaps, other AGNs may be produced by such a mechanism.

\end{abstract}





\section{Introduction}
\label{sec:introduction}

Many active galactic nuclei (AGNs) occasionally exhibit rapidly variable TeV emission, the origin of which is yet unknown. 
A notable example is the large amplitude, rapid TeV flares observed in M87 \citep{2006Sci...314.1424A,2012ApJ...746..151A}. 
The short duration of these extreme flares, roughly  the dynamical time of the central black hole (BH),
as well as other indicators (e.g., coincidence with core radio emission) suggest that the gamma-ray emission
originates from the very inner regions of the magnetosphere.

It has been proposed that charge starved magnetospheric regions (spark gaps) might provide sites for 
variable TeV emission \citep{2000PhRvL..85..912L,2011ApJ...730..123L,2016ApJ...833..142H,2017ApJ...845...77H,2017PhRvD..96l3006L}.   
Such spark gaps are expected to form when direct plasma supply
to the inner region of an  active magnetosphere, where jets are formed, is too low to screen out 
the magnetosphere.  Particle acceleration by the gap electric field then gives rise to copious  pair creation 
through the interaction of particles accelerated in the gap with soft photons emitted by the accretion flow.
Recent one-dimensional (1D) general relativistic Particle-in-Cell (GRPIC) simulations of local spark gaps \citep{2018A&A...616A.184L,2020ApJ...895..121C,2020ApJ...902...80K} 
and global 2D GRPIC simulations of active Kerr BH magnetospheres \citep{2019PhRvL.122c5101P,2020PhRvL.124n5101C,2021A&A...650A.163C}
indicate that the spark process is highly intermittent, conceivably inherently cyclic under certain conditions. 
However, the amplitude of the variations and the overall luminosity were found to depend rather sensitively on the pair creation 
opacity contributed by the soft radiation emitted from the surrounding accretion flow.  These simulations 
invoke steady external conditions, particularly the soft photon intensity that is given as input.  

In reality, it is likely that the accretion flow will undergo temporal variations that induce intermittencies in
the pair production opacity and the strength (and geometry) of the magnetic field advected by the accretion flow. 
It is anticipated that such external changes might lead to a nonlinear response of the gap and its TeV emission, e.g., by the development of the electric field due to the change of the optical depth and the electric current density, and the screening of that electric field by the created pairs thus created. 
The variations in the gap emission should be delayed with respect to the variations of the 
target photon intensity.  In principle, it might be possible to map such reverberations and extract information 
about the structure of the magnetosphere and the spark process, if the delays can be quantified and general relativistic time travel effects are 
taken into account. 

In this paper we study the response of a local spark gap to external changes using 1D GRPIC simulations.
We explore the gap dynamics induced by the change of different parameters, particularly
target photon spectrum and luminosity, and the global magnetospheric current. We find that the gap response is highly 
non-linear by virtue of Klein-Nishina (KN) effects and the sensitive dependence of the curvature radiation power on the energy of 
emitting particles, and that such changes can lead to production of strong flares of curvature TeV emission by the gap accelerating particles.

\section{The gap model}
\label{sec:model}
Our analysis is restricted to the response of a 1D, local spark gap.  The tacit assumption underlying the local 
gap model is that its activity does not affect significantly the global magnetospheric structure.  
The global electric current and the characteristics of soft-photon intensity are treated as input parameters in this model.
The calculations are performed using the 1D GRPIC code described in \citet{2018A&A...616A.184L} and \citet{2020ApJ...902...80K}. 
The dynamical equations are solved in a background Kerr spacetime, given in Boyer-Lindquist coordinates with the radial coordinate replaced by the tortoise coordinate $\xi$ to avoid the singularity on the horizon (see \citealt{2018A&A...616A.184L} for details). 

The computational domain extends from a radius of $\sim1.5r_g$ to $\sim4.3 r_g$, where $r_g=GMc^{-2}=1.5\times10^{14}M_9~{\rm cm}$ is the gravitational radius, $G$ is the Gravitational constant, $c$ is light speed, and $M\equiv10^9M_{\odot}M_9$ is a BH mass. For the results presented below we used a BH mass of $M=10^9M_\odot$ and a dimensionless spin parameter $a_{\ast}=0.9$.  The corresponding angular velocity of the BH is $\omega_{\rm H}=a_{\ast}c/2r_{\rm H}\approx 10^{-4} M_9^{-1}$~Hz, where $r_{\rm H}=r_g(1+\sqrt{1-a_{\ast}^2})$ is the horizon radius. 
For the angular velocity of the magnetic surface along which the gap is located we adopt the value $\Omega=\omega_{\rm H}/2$, such that 
the null charge surface, on which the Goldreich-Julian (GJ) charge density vanishes, is located at $r_{\rm null}=2r_g$.  For all cases
considered below we use a split monopole magnetic field configuration with magnetic field strength $B_{\rm H}/2\pi=10^3$ G on the horizon, 
and adopt an inclination angle $\theta=30^\circ$ (with respect to the BH rotation axis) for the flux surface that contains the gap.
An open boundary condition is imposed on both sides (boundaries) of the simulation box.
The number of grid cells used in a typical simulation is 32768, sufficient to resolve the skin depth, 
and the initial particle per cell number (PPC) is 45, which guarantees convergence (see \citealt{2020ApJ...902...80K} for details).
Finally, the time step is set by the Courant-Friedrichs-Lewy condition at the inner boundary.

\subsection{Emission and pair creation}

Two processes dominate the emission by particles accelerated in the gap, inverse Compton (IC) scattering and curvature emission. 
In essentially all cases considered here, the mean energy of curvature photons is considerably smaller than that of IC scattered photons,
rendering pair creation by curvature photons negligible, even in cases where the gamma ray luminosity is dominated by curvature emission
(see \citealt{2020ApJ...902...80K} for detailed estimates).   We therefore treat these two processes in a distinct manner. 
The IC scattered photons are included as charge-neutral particles in the simulation and their momenta are computed by solving 
the appropriate equations of motions.  The curvature emission, on the other hand, is computed at each time step using analytic formula
for single particle emission, and summing the contributions from all particles in the simulation box, taking into account time travel effects \citep{2018A&A...616A.184L,2020ApJ...902...80K}.  Pair creation is contributed solely by the interaction of IC photons with the background
soft photons, envisioned to emerge from the surrounding accretion flow.

IC scattering and pair creation are computed using Monte-Carlo methods that are incorporated in the PIC code, as described in \citet{2018A&A...616A.184L}.  The full KN cross section is used for both processes.  The background photon intensity
is taken, for simplicity, to be isotropic and homogeneous, with a power-law spectrum:
\begin{eqnarray}
I(\epsilon)=I_0(\epsilon/\epsilon_{\min})^{-p}, ~~\epsilon_{\min}\le\epsilon\le\epsilon_{\max}, 
\end{eqnarray}
where $\epsilon$ is the photon energy normalized by an electron rest mass energy, $m_{\rm e}c^2$. 
It is characterized by three key parameters: the minimum cutoff energy $\epsilon_{\min}$, the slope $p$, and the normalization 
$I_0$ that we, alternatively, measure by the fiducial optical depth 
\begin{eqnarray}
\tau_0=4\pi r_g\sigma_{\rm T}I_0/hc,
\end{eqnarray}
here $\sigma_{\rm T}$ is the Thomson cross section and $h$ is the Planck constant.  As in \citet{2020ApJ...902...80K}, to reduce
the computational cost, we ignore the IC scattering for the particles with $\gamma<10^7$, since at these energies the created 
photons are  below the pair production threshold.

\subsection{Dichotomic gap behaviour}

In \citet{2020ApJ...902...80K} we identified a dichotomy in the gap dynamics that depends on the sign of the magnetospheric current $j_0$, 
henceforth normalized by $[\omega_{\rm H}B(1+a_{\ast}^2)\cos\theta/2\pi]$. 
From this definition, the current carried by the outflowing plasma particles with GJ charge density at large distance is $j_0=-1.0$. For negative values, $j_0<0$, the gap exhibits a cyclic discharge, with superposed 
plasma oscillations, confined to the region around the null surface \citep{2020ApJ...902...80K}. The duty cycle and amplitude of these oscillations 
depend quite sensitively on the fiducial optical depth $\tau_0$ and minimum energy $\epsilon_{\min}$.  For positive currents 
we observed mainly rapid oscillations, with occasional opening of the gap at the boundaries in certain cases. 

Despite the inherent intermittency of the discharge dynamics, the long-term behaviour has been found to be quasi-regular.
This is not surprising given that all key parameters were fixed to constant values.  However, the high nonlinearity 
of the gap dynamics suggests that any temporal changes in any of these parameters should lead to a more chaotic, vigorous response.
This is the main motivation for the analysis described in the following section.  Since
negative current is anticipated in the polar jet section we restrict our calculations to this case.

\subsection{Simplistic modelling of external changes}

Recent GR magnetohydrodynamic (MHD) simulations (e.g., \citealt{Porth2019,Philippov2020,Dexter2020,nathanail20,Chashkina21}) indicate a complex accretion dynamics that depends on the initial setup, particularly
the initial magnetic field configuration.  At sufficiently high resolution, drastic changes in the accretion rate and
the structure of the inner magnetosphere have been observed during states of magnetically arrested disk, even for a simple initial magnetic field configuration.
The flaring behaviour appears to be regulated by a reconnecting equatorial current sheet \citep{Philippov2020,Chashkina21}, and if small scale field is 
initially present also by sporadic reconnection at the boundaries of interacting flux tubes with opposite polarities \citep{nathanail20,Chashkina21}. 
Formation of an equatorial current sheet with a feedback on the global magnetosphere has been observed also in recent global GRPIC simulations 
of active black holes  \citep{2021A&A...650A.163C}.  It is quite likely, therefore, that this complex behaviour will give rise to
changes in disk emission and magnetospheric currents over dynamical timescales. 

The radio observations indicate that the spectral peak flux of M87 would come from the inner region, $r < 7 r_g$ (Fig 16 of \citet{2021ApJ...911L..11E}). The peak frequency is consistent with $\sim 10^{-9} m_{\rm e}c^2$ used as $\epsilon_{\min}$ in our model. 
Then, the soft photon field may be sensitive to the accretion rate and have a short variability, $\lesssim10 r_g/c$. 

\begin{figure*}
 \begin{center}
  \includegraphics[width=160mm]{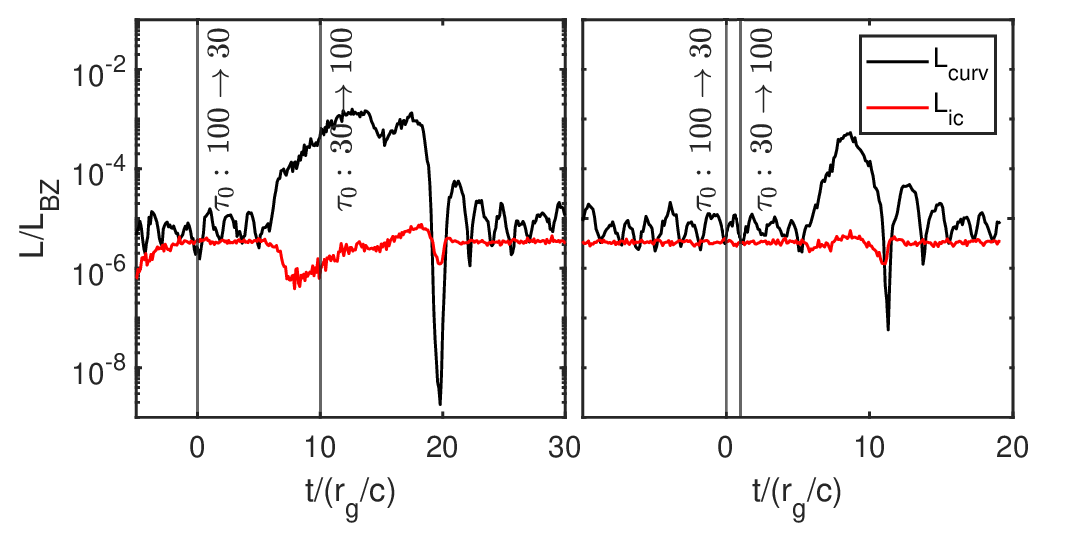}
   \caption{Light curves of curvature radiation (black) and IC emission (red) obtained at the outer boundary of the simulation box for a change of opacity
   from $\tau_0=100$ to $\tau_0=30$ at time $t= 0$ and back to the initial value at $t=10 t_g$ (left) and $t=1t_g$ (right).  The other model
   parameters remained fixed at the fiducial model values ($j_0=-1$, $\epsilon_{\min}=10^{-9}$ and $p=2$) throughout the entire event.
   All luminosities are normalized by the BZ luminosity (Equation \ref{eq:L_BZ}).}
  \label{fig:lightcurve0}
 \end{center}
\end{figure*}

Within the framework of our simplistic model this can be captured by allowing temporal changes in the model parameters, $j_0, \epsilon_{\min}, p, \tau_0$.  Our strategy is to study the gap response to abrupt changes in these parameters.  For our reference model we invoke the parameters 
$\tau_0=100$, $\epsilon_{\min}=10^{-9}$, $p=2$, and $j_0=-1.0$.  In each numerical experiment we change, instantaneously, one of those parameters
relative to the reference model, at a transition time that we define as $t=0$ (except for Fig. \ref{fig:lightcurve0}).  
The various models are listed in Table \ref{tab:parameter}.

\begin{table}
 \caption{Simulation Model Parameters. }
 \begin{center}
  \begin{tabular}{ccccc}

\hline
 Model & $j_0$ & $\tau_0$ & $\epsilon_{\min}$ & $p$ \\ \hline
 A     & -1.0  & 100 $\rightarrow$ 30 & $10^{-9}$ & 2  \\
 B     & -1.0  & 100 $\rightarrow$ 300 & $10^{-9}$ & 2  \\
 C     & -1.0  & 100 & $10^{-9}$ & 2 $\rightarrow$ 3  \\
 D     & -1.0  & 100 & $10^{-9}$ & 2 $\rightarrow$ 1.5  \\
 E     & -1.0  & 100 & $10^{-9} \rightarrow 10^{-8}$ & 2  \\
 F     & -1.0  & 100 & $10^{-9} \rightarrow 10^{-10}$ & 2  \\
 G     & -1.0 $\rightarrow$ -2.0  & 100 & $10^{-9}$ & 2  \\
 H     & -1.0 $\rightarrow$ -0.5  & 100 & $10^{-9}$ & 2  \\
 I     & -1.0 $\rightarrow$ -0.3  & 100 & $10^{-9}$ & 2  \\
 J     & -1.0 $\rightarrow$ -2.0  & 30 & $10^{-9}$ & 2  \\
 K     & -1.0 $\rightarrow$ -0.5  & 30 & $10^{-9}$ & 2  \\ 
 L     & -1.0 $\rightarrow$ -0.3  & 30 & $10^{-9}$ & 2  \\ \hline

  \end{tabular}
 \end{center}
  \label{tab:parameter}
\end{table}

\section{Results}
\label{sec:results}

\subsection{Change of disk emission properties}

We consider first the gap response to changes in the disk emission for a constant magnetospheric current $j_0=-1$. 
All luminosities are normalized by the Blandford-Znajek (BZ) power \citep{1977MNRAS.179..433B}, 
\begin{eqnarray}\label{eq:L_BZ}
L_{\rm BZ}=\frac{1}{16}a_{\ast}^2B_{\rm H}^2r_{\rm H}^2c,
\end{eqnarray}
for the set of parameters ($a_\ast$, $r_g$, $B_H$) adopted in these simulations.
We generally find that relatively modest changes in the parameters of the target photon intensity lead to a highly nonlinear response of the gap,
that appears as large amplitude flares of curvature emission, in the regime of mild opacities ($\tau_0 \lesssim 300$). The
IC emission is far less sensitive.  This is a consequence of the sensitivity of the curvature power to the Lorentz factor of emitting
pairs (see appendix \ref{sec:semi-analytic} for a detailed explanation), and the fact that IC scattering is in the deep KN regime.

To illustrate a typical time sequence of events, we show in Fig. \ref{fig:lightcurve0} an example where the fiducial 
optical depth has changed from $\tau_0=100$ to $\tau_0=30$ (at $t=0$) for a period of $\Delta t=10\ t_g$ (left) and $\Delta t =1t_g$ (right),  where $t_g\equiv r_g/c$,
and then back to its initial value, keeping all other parameters fixed.  As seen, a flare dominated by curvature radiation is emitted  from the outer boundary of the simulation box, with a delay of about $7\ t_g$ in both cases, roughly the propagation time of photons from the gap to the boundary (including redshift effects).  
The detailed shape of the flare is clearly different in each case.  In both cases the curvature power rises sharply initially by a factor of $\sim10$, with a rise time of about $1t_g$.
 In the case on the left there is a further, slower increase by a similar factor over time of about $5t_g$, whereas the sharp increase continues in 
 the other case (since the opacity by this time has already switched back to $\tau_0=100$, the number of emitting particle increases due to pair creation).
The flare rise seen in both cases results from excessive emission by pairs accelerating in the growing gap electric field following the sudden decline in opacity at $t=0$.  The opening of the gap near the null surface commences almost immediately, followed by a growth
of the electric field over several $t_g$ due to the decline in the density of photons and pairs, owing to the reduced opacity.    
This timescale determines the rise time of the flare, whereas the delay between the sudden change in the opacity and the flare onset is determined by the propagation time of the curvature photons from the gap to the outer boundary. In the case on left there is a subsequent plateau phase which we think indicates the saturation of the pair cascade
before the opacity is switched back.  This is not seen in the other case, presumably because the cycle was too short for saturation to occur before the opacity is switched back.  Ultimately, there is a sharp drop by a large factor over a time of about $1-2 t_g$, causing a large undershoot followed by a fast recovery to the pre-flare luminosity.

We find similar behaviour in other cases. The conclusion is that for abrupt changes the flare rise time is comparable to the gap size, although the exact shape of the light curve depends on details. We stress that
our calculations do not take into account multi-dimensional as well as general relativistic effects on the trajectories of the emitted photons. 
In reality, if the gap is located close enough to the horizon, these effects might dominate.  However, they can be computed accurately 
using ray tracing codes (e.g., \citealt{2021A&A...650A.163C}), so that the intrinsic timescales can be extracted.  It is also worth noting that gradual changes of the external
parameters (over times longer the the gap response time) are likely to smear out the intrinsic response time, but a delay is, nonetheless, 
expected.  The extreme TeV flares observed in M87 \citep{2006Sci...314.1424A, 2012ApJ...746..151A} and IC 310 \citep{2014Sci...346.1080A},
if produced by such a process, imply gap size of $\sim r_g$, and changes in external parameters over time scales not much longer than the light 
crossing time of the gap (see right panel in Fig. \ref{fig:lightcurve0}).

\begin{figure*}
 \begin{center}
  \includegraphics[width=200mm]{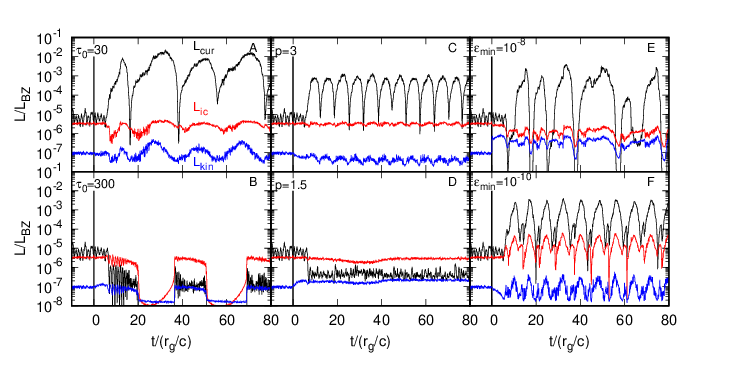}
   \caption{Light curves of curvature radiation (black), IC emission (red), and particle kinetic luminosity (blue) at the outer boundary of the simulation box. Transition between states occurs at $t=0$ (vertical lines) in all cases shown. In each panel one of the fiducial model parameters is changed to the indicated value, while the other parameters are kept fixed. All luminosities are normalized by the BZ luminosity.}
  \label{fig:lightcurve1}
 \end{center}
\end{figure*}

\begin{figure}
 \begin{center}
  \includegraphics[width=85mm]{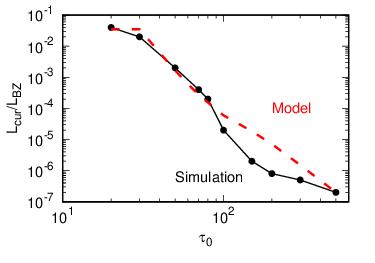}
   \caption{Curvature luminosity as a function of the fiducial optical depth $\tau_0$ (black line), compared with  the analytic estimate derived in appendix \ref{sec:semi-analytic} (dashed red line). }
  \label{fig:luminosity}
 \end{center}
\end{figure}

Figure \ref{fig:lightcurve1} shows mosaic of lightcurves obtained for models A$-$F in table \ref{tab:parameter}.
The black, red, and blue lines correspond, respectively, to fluxes of curvature radiation, IC emission and particles kinetic energy
at the outer boundary of the computation box.  In these runs, we start with the fiducial model and then, at time $t=0$ in the plots,
change one of the parameters abruptly and let the system relax to the new state.  As in Fig. \ref{fig:lightcurve0}, we observe a delay of 
about $7 t_g$ in the gap response in all cases.  After the delay, the activity quickly ($\sim t_g$) reaches another quasi-periodic state for 
the new parameter set, which is consistent, in all cases, with the results reported in \citet{2020ApJ...902...80K}. 
As seen, strong flares are produced in episodes in which the opacity effectively declines. 
This applies to the cases displayed in the three upper panels.  Steepening of the spectrum (increased $p$) results in a reduced 
optical depth due to reduction in the number of photons above the peak (i.e., above $\epsilon_{\min}$) that interact with lower energy pairs.
As for the change in $\epsilon_{\min}$, the response arises since the particle kinetic energy flux increases following the transition. 
In the fiducial case with $\epsilon_{\min}=10^{-9}$, 
particles lose their energy near the outer boundary by virtue of IC scattering in the Thomson regime. 
The typical energy of particles near the outer boundary is $\gamma\sim10^8$. 
Then, after switching to $\epsilon_{\min}=10^{-8}$, the same particles scatter in the KN regime,
thereby cool much slower.  As a result, their mean energy increase leading to significant enhancement in 
curvature emission.  

In general, we find that the highly nonlinear response we observed in many runs is
a combination of KN effects and the sensitive dependence of the curvature emission power on the energy of emitting particles. 
The dependence of the peak luminosity of curvature emission on the optical depth $\tau_0$ is delineated by the black line in Figure \ref{fig:luminosity}. 
The red line corresponds to the analytic result derived in appendix \ref{sec:semi-analytic}.
This sensitivity of $L_{\rm cur}$ to changes in opacity is the key factor that leads to the flaring behaviour seen in 
Figs. \ref{fig:lightcurve0} and \ref{fig:lightcurve1}.

\subsection{Change of magnetospheric current}

\begin{figure*}
 \begin{center}
  \includegraphics[width=200mm]{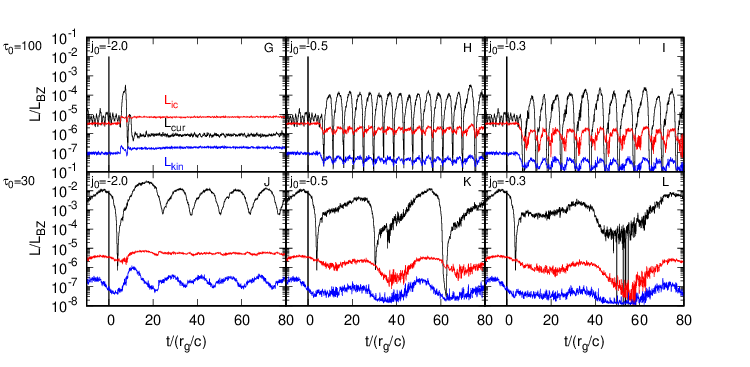}
   \caption{Light curves of curvature radiation (black), IC emission (red), and particle kinetic luminosity (blue) at the outer boundary of the simulation box. The fiducial optical depth is $\tau_0=100$ for upper panels and $\tau_0=30$ for lower panels. At $t=0$ (vertical lines), the magnetospheric current density changes from $j_0=-1.0$ to $j_0=-2.0$ (left), $j_0=-0.5$ (middle), or $j_0=-0.3$ (right). All luminosities are normalized by the BZ power.}
  \label{fig:lightcurve2}
 \end{center}
\vspace{1mm}
\end{figure*}


In the second set of numerical experiments, models G$-$L in table \ref{tab:parameter}, we 
keep the soft-photon intensity fixed and change the global current $j_0$ at time $t=0$.   Since the 
net electric current in the dynamical equation that governs the evolution of the gap electric field
is $(j_e - j_0)$, where $j_e$ is the current carried by pairs produce in the gap, changes in $j_0$ 
lead to a change in the net current and, consequently, an inductive response. 
For example, since the number density of particles in the gap is comparable to the GJ value times $|j_0|$, when $|j_0|$ increases (as in case G, where $j_0 = -1.0 \to -2.0$), the charge density in the gap is insufficient for compensating the current difference  $j_e-j_0$. As a result, the flux of curvature photons temporarily increases, as seen in the left panels in Fig. \ref{fig:lightcurve2}.

Figure \ref{fig:lightcurve2} shows the resultant light curves.  The upper three panels correspond to 
models G$-$I in table \ref{tab:parameter} ($\tau_0=100$) and the lower panels to Models J$-$L ($\tau_0=30$).  In all cases
shown the global current changes from $j_0=-1$ to the value indicated in the top label.
As seen, the curvature emission flux appears to be quite sensitive to $j_0$ for $\tau_0=100$, less so for the 
lower optical depth run.  Nonetheless, an increase by a factor of about 3 in the flux following the transition 
is observed also at $\tau_0=30$ (lower panels
in Fig. \ref{fig:lightcurve2}).  The reason for that trend is not entirely clear to us.  In the run with $\tau_0=100$ 
we observed significantly larger amplitudes of the electric field oscillations at lower values  of $j_0$, whereas 
for $\tau_0=30$ these oscillations appears much less sensitive to $j_0$.  We suspect that KN effects might play a role
in dictating the dependence of gap opening on the global magnetospheric current.

\section{Discussion}
\label{sec:discussion}

We studied the response of a black hole spark gap to external changes by means of 1D GRPIC simulations. We focused on
changes in the global magnetospheric current, and in the intensity of soft disk photons that dominate the pair-production and inverse-Compton opacities.
Such changes likely arise from the intermittency of the accretion process, as observed in recent GRMHD simulations (e.g., \citealt{Porth2019,Dexter2020,Philippov2020,nathanail20,Porth2021,Chashkina21}),
and conceivably the presence of small scale magnetic fields in the accretion flow \citep{nathanail20,Chashkina21}.  
Our analysis is restricted to local 1D gaps and neglects the feedback on the global magnetosphere.   

We generally find that in the regime of modest pair-production opacity, as anticipated in, e.g., M87 and possibly other AGNs,
sudden changes in the pair-production optical depth inside the gap give rise to 
strong flares of curvature TeV photons, with an intrinsic rise time of the order of the gap light crossing time and a similar delay between
the variation of the external radiation intensity and the onset of the flare. Changes in the global magnetospheric current 
can also lead to flaring episodes under appropriate conditions.   The highly nonlinear response of the gap stems from KN effects 
and the sensitive dependence of the curvature emission power on the energy of emitting particles.

Our analysis ignores time travel 
and general relativistic effects.  In reality, if the gap is located close to the horizon, one expects 
strong lensing to significantly affect the lightcurve.   
The extreme TeV flares observed in M87 \citep{2006Sci...314.1424A, 2012ApJ...746..151A} and IC 310 \citep{2014Sci...346.1080A},
if produced by such a process, imply gap size of $\sim r_g$, and similar changes in disk emission.  Alternatively, those flares 
can be produced by reconnection episodes in an equatorial current sheet near the horizon, hadronic process in the accretion disk \citep{2020ApJ...905..178K}, or cloud-jet interaction \citep{barkov12}.

\acknowledgments
We are grateful to the anonymous referee for constructive comments.
Numerical computations were performed on Cray XC50 at cfca of National Astronomical Observatory of Japan, 
and on Cray XC40 and Yukawa-21 at Yukawa Institute for Theoretical Physics, Kyoto University. 
This work was supported by JSPS Grants-in-Aid for Scientific Research 18H01245 (K.T.), 18H01246, 19K14712 and 21H01078 (S.K.).

\appendix
\section{Analytical estimate of the curvature luminosity}
\label{sec:semi-analytic}
To elucidate the sensitive dependence of the luminosity of curvature radiation on the model parameters, we calculate the luminosity using a simple analytic 
model of the gap.  The analytic results are compared with the curvature luminosity obtained from the simulations (Fig. \ref{fig:luminosity}).
The analytic model assumes that the curvature luminosity emitted from the outer gap boundary is dominated by emission of particles 
accelerated in the open gap region around the null charge surface, where the electric field $E_{||}(r)$ is strongest and oscillations can be neglected.
We ignore any velocity spread, as it is negligibly small, and suppose that at any given radius $r$ the energy of an emitting particle, $\gamma(r)$, is 
limited by curvature losses (as seen in Figs. \ref{fig:lightcurve0},\ref{fig:lightcurve1} and \ref{fig:lightcurve2}, IC losses are negligible in most cases of interest).   It is obtained by equating the curvature power of a single particle,
\begin{eqnarray}\label{eq:P_cur}
P_{\rm cur}(\gamma)=\frac{2}{3}\frac{e^2\gamma^4 c}{R_{\rm c}^2},
\end{eqnarray}
with the rate of energy gain by the electric force, $e c E_{||}(r)$:
\begin{eqnarray}\label{eq:gamma}
\gamma(r) =\left(\frac{3}{2}\frac{E_{\parallel}(r) R_{\rm c}^2}{e}\right)^{1/4}.
\end{eqnarray}
Here $e$ is a charge of an electron and $R_c$ is the curvature radius of the magnetic field line confining the accelerating particles, assumed to be constant throughout the gap. 
The total curvature luminosity is obtained by integrating $P_{\rm cur}$ over all the particles inside the gap:
\begin{eqnarray}\label{eq:L_cur}
L_{\rm cur}=2\pi\int_{r_{\rm in}}^{r_{\rm out}}\alpha^2(r)P_{\rm cur}(\gamma)nr^2dr, 
\end{eqnarray}
where $r_{\rm out}$ and $r_{\rm in}$ are the outer and inner gap boundaries, respectively, $\alpha(r)$ is the lapse function, and $n$ 
is the number density of emitting particles, taken to be equal to $|j_0|\omega_{\rm H}B(1+a^2_{\ast})\cos\theta/2\pi ec$ at $r=r_{\rm null}$. 

The electric field is obtained from Gauss' law, noting that 
inside the gap the net charge density of the counter-streaming electrons and positrons is well below the GJ density, $|\rho_e|\ll|\rho_{\rm GJ}|$.
With $\rho_{\rm GJ}(r)\propto[2(r/r_{\rm H})^{-2.5}-1]$ used as an approximate fit to 
the GJ charge density around the null charge surface, we thus have
\begin{eqnarray}\label{eq:E_para}
E_{\parallel}(r)=-4\pi\int_r^{r_{\rm out}}\rho_{\rm GJ}(r')dr'.
\end{eqnarray}
The radius of the outer gap boundary, $r_{\rm out}$, is treated as a free parameter.
For a given $r_{\rm out}$, we calculate the density of newly created pairs, as explained below, and use it to find
the inner boundary of the gap, $r_{\rm in}$, which is taken to be at the radius at which  the number density of created 
pairs is equal to the number density of particles injected into the gap.
We then iteratively seek the radii $r_{\rm out}$ and $r_{\rm in}$ that satisfy $(r_{\rm out}+r_{\rm in})/2=r_{\rm null}$.

To estimate the density of created pairs, we only consider the annihilation of IC gamma-ray photons with soft disk photons. 
The number density of scattered IC photons in an interval $\Delta r$ is 
\begin{eqnarray}\label{eq:n_gamma}
\Delta n_{\gamma}\sim\frac{\Delta r}{l_{\rm IC}}n, 
\end{eqnarray}
where $l_{\rm IC}$ is the mean free path for IC scattering. 
Since the scattering is not in the deep KN regime, we simply use the following approximation for the mean free path:
\begin{eqnarray}\label{eq:ell_ic}
l_{\rm IC}\sim\left\{ \begin{array}{ll}
r_g/\tau_0 & ~~~(\gamma\epsilon_{\min}\le1) \\
r_g\epsilon_{\min}\gamma/\tau_0 & ~~~(\gamma\epsilon_{\min}>1). \\
\end{array} \right.
\end{eqnarray}
The energy of a scattered photon satisfies
\begin{eqnarray}\label{eq:epsilon_ic}
\epsilon_{\rm IC}\sim\left\{ \begin{array}{ll}
0.1\gamma^2\epsilon_{\min} & ~~~(\gamma\epsilon_{\min}\le1) \\
0.1\gamma & ~~~(\gamma\epsilon_{\min}>1). \\
\end{array} \right.
\end{eqnarray}
We assume that a scattered photon is annihilated after propagating a distance comparable to the characteristic mean free path for the pair creation, 
\begin{eqnarray}\label{eq:ell_gamma}
l_{\gamma\gamma}\sim10\frac{r_g}{\tau_0}\left(\frac{\epsilon_{\rm tar}}{\epsilon_{\min}}\right)^{p}\epsilon_{\rm IC}\epsilon_{\rm tar},
\end{eqnarray}
where $\epsilon_{\rm tar}$ is the minimum energy of target photons, defined as 
\begin{eqnarray}\label{eq:epsilon_tar}
\epsilon_{\rm tar}=\max\{\epsilon_{\min},\epsilon_{\rm IC}^{-1}\}.
\end{eqnarray}
Using these estimates we integrate the transfer equation to obtain the number density of newly created pairs, and use it to 
determine the inner boundary of the gap, $r_{\rm in}$, assumed to be located where the integrated number density reaches the GJ value. 

As a rough estimate, using equations (\ref{eq:P_cur}) and (\ref{eq:L_cur}), the curvature luminosity is $L_{\rm cur}\propto\gamma^4 r_{\rm gap}$, where $r_{\rm gap}=r_{\rm out}-r_{\rm in}$ is the gap width. Using equations (\ref{eq:gamma}) and (\ref{eq:E_para}), the Lorentz factor of the particles is $\gamma\propto E_{\parallel}^{1/4}\propto r_{\rm gap}^{1/4}$. The gap width is roughly the minimum of the sum of two mean free path, $l_{\rm IC}$ and $l_{\gamma\gamma}$. In the KN regime, both mean free path scales as $\gamma/\tau_0$ from equations (\ref{eq:ell_ic}) and (\ref{eq:ell_gamma}). Then, the curvature luminosity scales as $L_{\rm cur}\propto r_{\rm gap}^2\propto \tau_0^{-8/3}$. 

Figure \ref{fig:luminosity} shows the calculated curvature luminosity as a function of the optical depth $\tau_0$ (red dashed curve), 
for $j_0=-1.0, \epsilon_{\min}=10^{-9}$, and $p=2$.  The analytic result is compared with the luminosity obtained 
from the numerical simulations.  As seen, the agreement is extremely good. The curvature luminosity in the analytic result scales as $L_{\rm cur}\propto \tau_0^{-3}$, roughly consistent with the above rough estimate. 

\bibliographystyle{apj_8}
\bibliography{ref}

\end{document}